\documentclass[]{aa}
\usepackage{amssymb,graphicx,natbib,txfonts}
\renewcommand{\d}{\mathrm{d}}

\authorrunning{M. Cacciato et al.}
\titlerunning{Combining weak and strong lensing in cluster potential
  reconstruction}

\begin{document}

\title{Combining weak and strong lensing in cluster potential
  reconstruction}

\author
  {Marcello Cacciato\inst{1,3}, Matthias Bartelmann\inst{2}, Massimo
   Meneghetti\inst{2}, Lauro Moscardini\inst{1,4}
   \institute
     {$^1$ Dipartimento di Astronomia, Universit\`a di Bologna, 
      via Ranzani 1, 40127 Bologna, Italy\\
      $^2$ Zentrum f\"ur Astronomie, ITA, Universit\"at Heidelberg,
      Albert-\"Uberle-Str.~2, 69120 Heidelberg, Germany\\
      $^3$ Max-Planck-Institut f\"ur Astronomie, K\"onigstuhl ~17,
      69117 Heidelberg, Germany\\
      $^4$ INFN, Sezione di Bologna, Italy}}

\date{\emph{Astronomy \& Astrophysics, in press}}

\abstract{We propose a method for recovering the two-dimensional
  gravitational potential of galaxy clusters which combines data from
  weak and strong gravitational lensing. A first estimate of the
  potential from weak lensing is improved at the approximate locations
  of critical curves. The method can be fully linearised and does not
  rely on the existence and identification of multiple images. We use
  simulations to show that it recovers the surface-mass density
  profiles and distributions very accurately,even if critical curves
  are only partially known and if their location is realistically
  uncertain. We further describe how arcs at different redshifts can
  be combined, and how deviations from weak lensing can be included.}

\maketitle 

\section{Introduction}

Weak gravitational lensing constrains the projected mass distributions
of galaxy clusters with an angular resolution of $\gtrsim0.5'$, while
strong lensing occurs typically not farther than $(0.5-1)'$ from
cluster cores. Nonetheless, both phenomena are due to the same
gravitational potential. Recent observations of galaxy clusters with
the ACS camera on-board HST (see \citealt{BR05.4} for an example) have
revealed large numbers of arcs in individual galaxy clusters, for
which Abell~1689 is an outstanding example. In such clusters, strongly
lensed images provide numerous constraints on the lensing potential,
and the question is raised how weak and strong lensing are best
combined in joint reconstructions of the lensing mass distribution.

Several methods were recently proposed which rely on the
identification of multiply-imaged systems
\citep{BR05.2,BR05.1,DI05.1}. We propose an alternative method here
for which multiple images are not necessary and thus do not need to be
identified. Based on a least-$\chi^2$ minimisation, a first estimate
of the lensing potential is obtained from weak lensing alone,
following the approach suggested by \cite{BA96.3} and
\cite{SE98.4}. Where the position of critical curves can be
approximately identified due to the presence of strongly-lensed
images, this estimate is improved by requiring that the determinant of
the Jacobian matrix vanish. The method is easily generalised to
sources at different redshifts. It can be fully linearised and allows
deviations from the strict weak-lensing limit to be accounted for by
simple iteration.

We outline the main ideas underlying the method in Sect.~2. After
reviewing the necessary formalism of gravitational lensing in Sect.~3,
we describe the method in Sect.~4 and illustrate its feasibility using
numerical simulations in Sect.~5. Conclusions are given in Sect.~6,
and technical detail is summarised in the Appendix.

\section{Overview}

The method we are proposing rests on two central ideas. First, both
weak and strong lensing must be described by the same underlying,
two-dimensional lensing potential. Data from weak and strong lensing
observations must thus be expressed in terms of constraints on the
potential. Second, strong lensing can be located in several ways, of
which the identification of multiple images is one, and the existence
of critical curves is another. Recent images of strongly-lensing
galaxy clusters illustrate that gravitational arcs in many cases allow
an accurate, piece-wise localisation of critical curves. On the other
hand, multiple images need to be identified first and are only caused
by sources close to and inside caustics, while sources outside
caustics can still be used to identify critical curves.

Based on these ideas, our algorithm can be summarised very simply: we
wish to find a map of the lensing potential, discretised on a grid,
which is determined such as to optimise the agreement with the
observed data. The gravitational shear caused by weak lensing imprints
coherent distortions on the images of background galaxies. To
lowest-order approximation, their average ellipticity is an unbiased
estimator of the local shear. The lensing potential must be arranged
such as to reproduce this measured shear. Critical curves occur at
singular points of the lens mapping, which locally constrain the
curvature matrix of the potential. Thus, our algorithm aims at
optimally constraining the potential such that its tidal field
reproduces the measured shear, and its curvature reproduces the
critical curves.

Several complications come to mind immediately. First, there is a
problem of scales. Given number densities of $\sim30-40$ background
galaxies per square arc minute, the angular resolution of weak-lensing
observations is limited to $\gtrsim0.5'$. Strong lensing, however,
typically happens within $\lesssim1'$ of cluster centres. Both can be
combined only if grids of high resolution in the core and low
resolution farther outside are used. For the practical implementation,
we choose to first reconstruct the potential at low resolution on a
coarse grid, then refine the grid everywhere in the core area where
critical curves exist, fill the refined grid by suitable interpolation
from the coarse grid, and correct the potential there such as to
reproduce the critical curves.

Second, galaxy ellipticities reflect the shear $\gamma$ only in the
limit of weak lensing, $\kappa\ll1$. Deviations occur towards critical
curves. Outside critical curves, ellipticities constrain the
\emph{reduced shear} $\gamma/(1-\kappa)$, and its inverse complex
conjugate $1/g^*$ inside. As other studies (e.g.~\cite{BR05.1}) have
remarked earlier, this is not a problem of principle because it can be
solved in a quickly converging iteration.

Third, arcs in clusters typically originate from sources at different
redshifts. Since the geometrical efficiency of a fixed lens increases
monotonically with source redshift, so does the lensing
potential. Arcs at different redshifts thus constrain the curvature of
a lensing potential with the same shape but different
amplitude. However, the potential grows with source redshift by a
linear factor. This suggests the introduction of a reference potential
for one arbitrary source redshift, to which the potentials for sources
at other redshifts can linearly be transformed.

Fourth, critical curves can almost never be traced throughout a
cluster, but only piece-wise. This is irrelevant for our purposes
since the constraint that the curvature of the lensing potential must
reproduce the critical curves is purely local. As the distortion of
arcs close to critical curves varies rapidly with position,
measurement errors for the location of critical curves are typically
small. Alternatively, constraints on critical curves obtained from
parameterised lens models can be used in a hybrid approach.

Finally, all constraints can be expressed in terms of second
derivatives of the lensing potential. In the limit of weak lensing,
the constraints are linear since the source ellipticity is then a
linear combination of second potential derivatives. Moving into the
strong-lensing regime, the ellipticity constraints can be kept linear
by means of an iterative procedure. We devise a scheme in which even
the constraints from critical curves can be expressed by a linear
correction term. While this linearity is not conceptually important,
it is practically because it allows minimisations by matrix
inversions.

\section{Basic formalism}

We start by reviewing the basic formalism for gravitational lensing as
we shall need it in the course of the paper. Dealing with isolated
lensing systems such as galaxy clusters, we adopt the thin-lens
approximation, according to which the lensing mass distribution is
projected onto a lens plane perpendicular to the
line-of-sight. Sources are located on source planes which are also
perpendicular to the line-of-sight. The lens system is characterised
by the three angular-diameter distances $D_\mathrm{l,s,ls}$ between
the observer and the lens, the observer and the source, and between
lens and source, respectively.

All relevant properties of the lens system are then contained in the
scalar lensing potential $\psi$, which is the suitably projected and
rescaled Newtonian gravitational potential $\Phi$ of the lens,
\begin{equation}
  \psi(\vec\theta)=\frac{2}{c^2}
  \frac{D_\mathrm{ls}}{D_\mathrm{l}D_\mathrm{s}}\,
  \int\Phi(D_\mathrm{l}\vec\theta)\,\d z
\label{eq:01}
\end{equation}
(e.g.~\citealt{SC92.1}). Obviously, the lensing potential depends on
the source redshift. Assuming that the lensing mass distribution is
the same for sources on different source planes, we can still
introduce a single scalar potential $\bar\psi(\vec\theta)$ for a
fiducial source redshift $\bar z_\mathrm{s}$, and then scale the
potential to other source redshifts $z_\mathrm{s}$ according to
\begin{equation}
  \psi(\vec\theta,z_\mathrm{s})=\bar\psi(\vec\theta)\,
  \frac{D_\mathrm{ls}(z_\mathrm{s})}{D_\mathrm{s}(z_\mathrm{s})}
  \frac{D_\mathrm{s}(\bar z_\mathrm{s})}
       {D_\mathrm{ls}(\bar z_\mathrm{s})}\;.
\label{eq:02}
\end{equation}
The linearity of this transformation allows the linear reconstruction
of the single, fiducial lensing potential $\bar\psi$ from sources on
multiple source planes.

The two-dimensional, projected mass distribution of the lens is
described by the convergence $\kappa$, which is the surface mass
density $\Sigma$ in units of
\begin{equation}
  \Sigma_\mathrm{cr}=\left(\frac{4\pi G}{c^2}
  \frac{D_\mathrm{l}D_\mathrm{ls}}{D_\mathrm{s}}\right)^{-1}\;.
\label{eq:04}
\end{equation}
Distortions caused by the lens are characterised by the trace-free,
symmetric shear tensor with the two components $\gamma_{1,2}$. Both
are linear combinations of second derivatives of $\psi$,
\begin{equation}
  \gamma_1=\frac{1}{2}\left(\psi,_{11}-\psi,_{22}\right)\;,\quad
  \gamma_2=\psi,_{12}\;,\quad
  \kappa=\frac{1}{2}\vec\nabla^2\psi\;.
\label{eq:03}
\end{equation}
The lens mapping relates the source position $\vec\beta$ to the image
position(s) $\vec\theta$,
\begin{equation}
  \vec\beta=\vec\theta-\vec\nabla\psi(\vec\theta)\;.
\label{eq:05}
\end{equation}

For sources which are small compared to typical scales of the
(reduced) deflection angle $\vec\alpha=\vec\nabla\psi$, the lens
mapping can be linearised. Imaging is then locally characterised by
the Jacobian matrix
\begin{equation}
  \mathcal{A}=\frac{\partial\vec\beta}{\partial\vec\theta}=
  \left(\begin{array}{cc}
    1-\kappa-\gamma_1 & -\gamma_2\\
    -\gamma_2 & 1-\kappa+\gamma_1\\
  \end{array}\right) \ .
\label{eq:06}
\end{equation}
This shows that $\kappa$ is responsible for shrinking or stretching
images isotropically, while $\gamma$ causes anisotropic deformation.

A sufficiently small circular source of radius $r$ is imaged as an
ellipse with semi-major and -minor axes $a=r(1-\kappa-\gamma)^{-1}$
and $b=r(1-\kappa+\gamma)^{-1}$, respectively. The ellipticity,
defined as $\epsilon\equiv(a-b)/(a+b)$, is thus
\begin{equation}
  \epsilon=\frac{\gamma}{1-\kappa}\equiv g\;,
\label{eq:07}
\end{equation}
defining the so-called \emph{reduced shear} $g$. In the weak-lensing
regime, characterised by $\kappa\ll1$ and $\gamma_{1,2}\ll1$, the
ellipticity approximates the shear, $\epsilon\approx\gamma$. In
gravitational lenses capable of strong lensing, the weak-lensing
approximation typically fails very close to the centre. While the
linear relation between ellipticity and shear can then be applied in
the outskirts of the lens, corrections may become necessary near the
core.

Critical curves $\vec\theta_\mathrm{c}$ are closed curves in the lens
plane consisting of points where $\mathcal{A}$ cannot be inverted,
$\det\mathcal{A}(\vec\theta_\mathrm{c})=0$. Their images in the source
plane under the lens mapping (\ref{eq:05}) are called caustics; they
are given by $\vec\beta_\mathrm{c}=\vec\theta_\mathrm{c}-
\vec\nabla\psi(\vec\theta_\mathrm{c})$. Obviously, the relation
\begin{equation}
  (1-\kappa)^2-|\gamma|^2=0
\label{eq:08}
\end{equation}
is satisfied along critical curves.

Any measurement of gravitational lensing which is based on local
distortion information alone cannot distinguish between lensing with
the Jacobian matrices $\mathcal{A}$ and $\lambda\mathcal{A}$, with
$\lambda\ne0$. The matrix $\lambda\mathcal{A}$ produces images which
are isotropically stretched by the factor $\lambda$, but with
identical ellipticity $\epsilon$ as the matrix $\mathcal{A}$. This
invariance against the transformation
$\mathcal{A}\to\lambda\mathcal{A}$ causes $\kappa$ and $\gamma$ to be
invariant against the transformation
\begin{equation}
  \kappa\to(1-\lambda)+\lambda\kappa\;,\quad
  \gamma\to\lambda\gamma\;,
\label{eq:09}
\end{equation}
which obviously leaves the reduced shear (\ref{eq:07}) invariant. This
invariance transformation was first described as a ``mass-sheet
degeneracy'' (\citealt{FA85.1,SC95.1}) because it tends to
$\kappa\to(1-\lambda)+\kappa$ in the limit of $|1-\lambda|\to0$.

We emphasise in the context of our joint reconstruction method that
the critical curves are also invariant under the transformation
(\ref{eq:09}). Obviously, the defining condition $\det\mathcal{A}=0$
for critical curves is unchanged if $\mathcal{A}$ is multiplied by
$\lambda\ne0$. Translated to the underlying lensing potential, the
transformation (\ref{eq:09}) allows transformations of the potential
of the form
\begin{equation}
  \psi\to\lambda\psi+\frac{1-\lambda}{2}(\theta_1^2+\theta_2^2)+
  a\theta_1+b\theta_2+c\;,
\label{eq:10}
\end{equation}
with arbitrary constants $a$, $b$ and $c$. We shall later take
advantage of this transformation for adjusting the lensing potential
obtained from weak- and strong-lensing constraints.

\section{Outline of the method}

We propose to combine local constraints from weak and strong lensing
on the lensing potential $\psi$ in the reconstruction of
two-dimensional lensing mass distributions. We aim at the lensing
potential because it is the smoothest lensing quantity available and
because it provides a complete description of all lensing phenomena at
least in the approximation of a single-lens plane. The lens is covered
by a grid. The values $\psi_i$ of the lensing potential in the grid
cells are considered as the model parameters and the method seeks to
find a set of potential values $\{\psi_i\}$ such that the agreement
between the data and the model is optimised in a least-$\chi^2$ sense.

We introduce a $\chi^2$ function which is the sum of two terms,
$\chi_\mathrm{w}^2$ and $\chi_\mathrm{s}^2$, which encode information
provided by weak and strong lensing, respectively. While
$\chi_\mathrm{w}^2$ is defined on a grid covering the cluster with low
resolution, $\chi_\mathrm{s}^2$ is restricted to a refined grid near
the cluster core. Both contributions will be defined in the following
subsections.
 
\subsection{Weak lensing}

Following the approach suggested by \cite{BA96.3} and \cite{SE98.4},
one choice for the weak-lensing contribution to the $\chi^2$ function
is
\begin{equation}
  \chi_\mathrm{w}^2=\sum_{i=1}^n
  \frac{|\epsilon_i-\hat\epsilon_i(\psi_j)|^2}{\sigma_{w i}^2}\;,
\label{eq:11}
\end{equation}
where $n$ is the number of grid cells covering the lens plane and
$\hat\epsilon_i$ is the expectation value for the ellipticity averaged
within the $i$th cell. As already stated, $\hat\epsilon_i$ depends on
second derivatives of the deflection potential $\psi$.

The number of cells, $n$, must be chosen so that the grid cells are
large enough for reasonably accurate measurements of the averaged
ellipticities $\epsilon_i$ , and yet small enough for the lensing
potential $\psi$ not to change appreciably across a grid cell. We
adopt the common assumption that intrinsic source ellipticities are
randomly oriented and thus tend to zero when averaged within
sufficiently large samples. The typical standard deviation of
intrinsic ellipticities from zero is $\sigma_\epsilon\approx0.3$
\citep{BR96.1}. Requiring that the noise of a shear measurement within
a grid cell due to the intrinsic ellipticities be at or below the
10\%-level, we need to locally average over a number $N$ of galaxies
determined by
\begin{equation}
  \frac{\sigma_\epsilon}{\sqrt{N}}\approx
  \frac{0.3}{\sqrt{N}}\lesssim0.1\;,
\label{eq:13}
\end{equation}
thus $N\gtrsim10$. At a surface density of, say,
$30\,\mathrm{arcmin}^{-2}$ for the background sources
(e.g.~\citealt{FO96.1,RI04.2}), this implies grid cells of
$\gtrsim35''$ side length. In the ``concordance'' $\Lambda$CDM
cosmology, $1\,h^{-1}\mathrm{Mpc}$ spans $\sim5'$ at redshift
$\sim0.3$. The virial diameter of a massive cluster at that redshift
is thus contained in fields of $\sim(10'\times10')$, approximately
corresponding to $\sim(20\times20)$ pixels. This illustrates the grid
resolution which we can expect weak-lensing cluster mass
reconstructions to achieve. Then, the typical noise level
$\sigma_{\mathrm{w}i}$ per pixel is $\approx0.1$, as follows from
(\ref{eq:13}).

The expectation value for ellipticities, $\hat\epsilon_i$, depends on
a combination of both the convergence and the shear,
\begin{equation}
  \hat\epsilon_i(\psi)\equiv\hat g=\left\{
    \begin{array}{lll}
      \frac{\gamma}{1-\kappa} &
      \mbox{where}\quad 1-\kappa-|\gamma|\ge0 & \quad\mbox{(a)}\\
      &&\\
      \frac{1-\kappa}{\gamma^*} &
      \mbox{elsewhere} & \quad\mbox{(b)}\\
    \end{array}
  \right.
\label{eq:expect_value}
\end{equation}
Expression (b) is restricted to the innermost region of the lens where
both the convergence and the shear must be comparably large.

Based on Eq.~(\ref{eq:expect_value}), the minimisation of
$\chi_\mathrm{w}^2$ in (\ref{eq:11}) yields a non-linear relation
between the measured ellipticities and the deflection potential. This
technical problem is conveniently solved by an iterative procedure
(see e.g.~\cite{BR05.1}). Starting from $\kappa_i^{(0)}=0$ on all grid
cells, subsequent iterative approximations $\{\psi_i^{(k)}\}$ to the
lensing potential are obtained by minimising
\begin{equation}
  \chi_\mathrm{w}^2=\sum_{i=1}^n
  \frac{1}{\sigma_{\mathrm{w}i}^2}\left|
    \epsilon_i-\hat g_i\left(\psi_j^{(k+1)},\kappa_i^{(k)}\right)
  \right|^2\;
\label{eq:12}
\end{equation}
where $\hat g$ has to be chosen from two cases identified in
(\ref{eq:expect_value}).

Although this iteration provides an adequate solution for the
potential, it may not be needed in actual reconstructions because of
the difference in scales between weak and strong lensing. Strong
lensing occurs typically within $30''$ to $1'$ from cluster cores,
while the resolution that can be achieved by weak lensing is
$\gtrsim35''$. Thus, strong lensing is confined to the innermost few
cells of the weak-lensing grid.

Once the potential values are found which reproduce the weak-lensing
data, strong-lensing constraints can be added as described in the next
subsection.

So far, we have essentially reviewed the cluster reconstruction
approach proposed by \cite{BA96.3} and extended by maximum-entropy
regularisation in \cite{SE98.4}. While it was suggested in
\cite{BA96.3} to minimise $\chi_\mathrm{w}^2$ with the
conjugate-gradient method, the finite differencing on the
lensing-potential grid needed to obtain expectations for $\gamma_i$
implies that, in the weak-lensing regime,
\begin{equation}
  \frac{\partial\chi_\mathrm{w}^2}{\partial\psi_i}=0
\label{eq:14}
\end{equation}
is a linear equation in the $\psi_i$ which can be solved using matrix
inversion. The $\psi_i$ are thus obtained from
\begin{equation}
  \psi_j=\mathcal{B}_{jk}^{-1}\mathcal{V}_k\;,
\label{eq:16}
\end{equation}
with a sparse matrix $\mathcal{B}$ and a data vector $\mathcal{V}$
which are detailed in the Appendix.

This was noted by \cite{BR05.2,BR05.1}, who have recently proposed and
applied an alternative algorithm for combining weak and strong cluster
lensing. The main difference between their and our algorithm is the
way how constraints from strong lensing are taken into account. While
\cite{BR05.1} use information from multiply-imaged sources, we
advocate constraining the potential using the (approximate) knowledge
of the critical curves from the location of large arcs, as described
in the next subsection.

\cite{DI05.1} recently proposed another cluster reconstruction
procedure which also combines weak and strong-lensing data. Their idea
is to expand the projected cluster mass distribution into a set of
basis functions which are then constrained individually using shear
and strong-lensing data. The minimisation then proceeds iteratively,
using an adaptive grid to cover the cluster field.

\subsection{Strong lensing}

Strong lensing in clusters gives rise to highly distorted large arcs
which occur in the immediate vicinity of critical curves in the lens
plane. There, by definition (\ref{eq:08}), the positions of large arcs
approximate the locations where $\det\mathcal{A}=0$. If written in
terms of the lensing potential $\psi$, this condition translates into
an expression which is quadratic in the second derivatives of $\psi$.

We suppose that a solution $\psi_j$ has already been obtained from
weak lensing using (\ref{eq:16}). It exists on a coarse grid adapted
to the low resolution of weak-lensing measurements. This grid is now
refined near the cluster core to a resolution adapted to the critical
curve(s). Since strong lensing typically occurs within the innermost
arc minute around the cluster core, this grid will be much finer and
smaller than the grid introduced for weak lensing. It thus forms a
sub-grid which improves the resolution of the few central cells of the
weak-lensing grid that contain the large arcs, and thus parts of the
critical curve.

In the idealised case of a fully known critical curve, we could
identify a continuous chain of sub-grid cells covering the critical
curve. Let the number of these cells be $n^*$, then the contribution
of strong lensing to the $\chi^2$ function could be
\begin{equation}
  \chi_\mathrm{s}^2=\sum_{i=1}^{n^*}\,
  \frac{(\det\mathcal{A})_i^2}{\sigma_{\mathrm{s}i}^2}=
  \sum_{i=1}^{n^*}\frac{\left[
    (1-\kappa)^2-|\gamma|^2
  \right]^2}{\sigma_{\mathrm{s}i}^2}\;,
\label{eq:150}
\end{equation}
expressing the expectation that the Jacobian determinant be zero
within the tolerance expressed by $\sigma_{\mathrm{s}i}$ in all
sub-grid pixels covering the critical curve.

The derivative of the above function with respect to the potential
values can be incorporated into the matrix approach as detailed in the
Appendix. This leads a term to be introduced to the previous system
(\ref{eq:16}),
\begin{equation}
  \mathcal{B}_{jk}\psi_k=\mathcal{V}_j-\mathcal{T}_j\;.
\end{equation}
The implementation of this approach is addressed in the next
subsection.

The resolution of the central sub-grid introduced for the
strong-lensing constraints has to be high enough to follow the
critical curve with sufficient accuracy. The tolerance
$\sigma_{\mathrm{s}i}$ quantifies tolerable deviations of
$(\det\mathcal{A})_i$ from zero. It combines the uncertainty in the
position of the critical curve, which must be estimated from the
observed arc positions, and the deviation of $\det\mathcal{A}$ from
zero expected within one sub-grid pixel.

The second contribution can be suppressed to a negligible level
because we are free to choose the sub-grid resolution. The first
contribution can be estimated by considering the deviation of
$\det\mathcal{A}$ from zero at distance $\delta\theta$ from the
critical curve,
\begin{equation}
  \sigma_\mathrm{s}\approx\left.
    \frac{\partial\det\mathcal{A}}{\partial\theta}
  \right|_{\theta_\mathrm{c}}\delta\theta\approx
  \frac{\delta\theta}{\theta_\mathrm{E}}\;,
\label{eq:16c}
\end{equation}
approximating the derivative of the Jacobian determinant at the
critical curve by the inverse of the Einstein angle
$\theta_\mathrm{E}$. This is exact for the tangential arcs formed by
an isothermal sphere, and reasonably accurate for similar lens
models. Assuming uncertainties in the positions of critical curves of
order $\delta\theta\approx1''$ and Einstein radii of order
$\approx30''$, we find $\sigma_\mathrm{s}\approx3\times10^{-2}$.

We believe that our approach has two distinct advantages compared to
methods using multiple images for constraining the lens model with
strong-lensing data. First, it can be used simply based on the
approximate knowledge of the location of the critical curves without
the need to have or identify multiple arcs which are due to the same
source. Prominent counter-arcs are often missing in strongly lensing
clusters, which reflects their lack of axial symmetry
\citep{GR88.1,GR89.1,KO89.1}. Thus, methods relying on the
identification of multiple images are only applicable to clusters
which produce multiple large arcs, and in which the multiple images
can be attributed to single sources with a high degree of
certainty. Second, methods based on multiple images require that all
images identified as belonging to the same source be imaged on a
single source. Even if an assumed model satisfies this requirement, it
must be tested whether the source would have any additional images
which are not observed. However, the inversion of lens models which is
necessary to search for all images of a source is a highly nonlinear
procedure. It thus appears to us that the combination of weak and
strong-lensing information by identifying critical curves (or parts
thereof) has substantial advantages compared to methods identifying
multiple images.

We emphasise that the method is fully local. Strong lensing
constraints can be imposed for individual grid points. Although
knowing the entire critical curves is desirable, only partial
knowledge is necessary. Of course, the proposed reconstruction
algorithm works best when the whole critical curve is
available. Critical curves can be estimated from parametrised lens
models. As an example of the feasibility of this procedure, we refer
to \cite{SA03.1} and \cite{ME05.1}. The authors show that observed arc
positions constrain the whole critical line.

In detail, \cite{ME05.1} perform a $\chi^2$ minimisation constraining
the critical curves that pass through the arcs. The reconstructed
critical curves fit the real ones with high accuracy (see Sect.~5 for
a quantitative comparison). It should be noted that such a method
depends on the ellipticity of the deflection potential. However, this
aspect is not a weakness in our case because the weak-lensing
reconstruction performed in the first step is able to supply that.

Arcs in clusters may appear at different redshifts and thus trace
different critical curves. Such information can be built into the
$\chi^2$ function by means of the redshift scaling (\ref{eq:02}). This
could be achieved by identifying one critical curve (or arc system) as
fiducial with a redshift $\bar z_\mathrm{s}$, and to scale the
critical curves for all other systems according to their
(spectroscopic or photometric) redshift by applying appropriate
distance factors to the lensing potential $\psi$.

\subsection{$\chi^2$ minimisation}

We now proceed to describe how the two contributions to $\chi^2$ are
joined. We begin by considering the weak-lensing constraint only,
i.e.~using the constraint
leads to
\begin{equation}
  \frac{\partial\chi^2}{\partial\psi_i}=0\quad\to\quad
  \frac{\partial\chi_\mathrm{w}^2}{\partial\psi_i}=0\;.
\end{equation}
This step may require an iteration if deviations from the weak-lensing
limit need to be taken into account.

The weak-lensing solution $\psi_i$ obtained in this way on the coarse
grid is improved on the fine grid as follows. Bi-cubic interpolation
is used on the reconstructed potential near the cluster core in order
to achieve a resolution high enough to incorporate the strong-lensing
constraints. We emphasise that the interpolation is carried out on the
lensing potential because it is the smoothest quantity available, and
that the interpolation scheme has to be of sufficiently high order for
the convergence and the shear to be continuous. In fact, since second
derivatives of the potential are needed, interpolation schemes up to
second order will fail.

The location and shape of the fine grid onto which the potential is
interpolated from the weak-lensing solution depends entirely on the
strong-lensing constraints. There can be single or multiple arc
systems which constrain the Jacobian determinant to zero at isolated
locations near the cluster cores, or approximations to the entire
critical curve may be available from strong-lensing reconstructions of
parametrised cluster mass models.

We interpolate the potential onto a square-shaped grid enclosing all
critical curves of our model cluster. This is not only the simplest
choice, but also driven by the idea that the central region needs to
be described in closer detail. More complicated adaptive grids could
be chosen to follow the critical curve(s) more specifically.

On the fine grid, the interpolated weak-lensing solution $\psi_i^*$ is
improved by requiring
\begin{equation}
  \frac{\partial\chi^2}{\partial\psi_i}=0 \quad\Rightarrow\quad
  \frac{\partial\chi_\mathrm{w}^2}{\partial\psi_i^*}=
  -\frac{\partial\chi_\mathrm{s}^2}{\partial\psi_i^*}\;,
\label{eq:17}
\end{equation}
which leads to the set of linear equations for $\psi_k^*$ on the fine
grid
\begin{equation}
  \mathcal{B}_{jk}\psi_k^*=\mathcal{V}_j-\mathcal{T}_{j^\prime}\;,
\label{eq:18}
\end{equation}
where $\mathcal{T}_{j^\prime}$ is a data vector containing the information on
the critical curves and the prime denotes that only few grid cells
in the inner cluster region are constrained. Details are shown in the
Appendix. If the interpolated weak-lensing solution already satisfies
$\det\mathcal{A}=0$ along the critical curve(s),
$\mathcal{T}^*_j=0$. Thus, $\mathcal{T}^*_j$ quantifies by how much
$\psi_k^*$ needs to be corrected due to the strong-lensing
constraints.

We point out again that the potential reconstructed from weak and
strong lensing alike allows transformations of the form (\ref{eq:10})
because neither the weakly-lensed ellipticities nor the critical
curves are changed by them. This allows smoothly matching the
potential values on the fine and coarse grids.

\section{Testing the method}

\subsection{Synthetic data}

In order to test and illustrate the proposed method, we apply it to
synthetic background images lensed by a simulated galaxy cluster taken
from the set described in \cite{BA98.2}. The cluster is simulated
within the ``concordance'' $\Lambda$CDM model with a mass resolution
of $1.0\times10^{10}\,h^{-1}M_\odot$. The current matter density
parameter is $\Omega_{\mathrm{m}0}=0.3$, the cosmological constant
is $\Omega_{\Lambda0}=0.7$ and the Hubble parameter is $h=0.7$. The
cluster's redshift is $z=0.35$, and its total mass is
$1.4\times10^{15}\,h^{-1}M_\odot$. It is embedded in a cube of
$5\,h^{-1}\mathrm{Mpc}$ (comoving) side length, corresponding to
$17.8$ arc minutes. At a coarse-grid resolution of $32\times32$
pixels, one pixel has a side length of $33''$, in good agreement with
the resolution constraint estimated for weak lensing above.

Lensed data are synthesised by randomly distributing elliptical
sources on the source plane, which we place at $z_\mathrm{s}=1$ for
simplicity. The sources have random intrinsic orientations, axis
ratios which are drawn randomly from the interval $[0.5,1]$, and axes
determined such that their area equals that of a circle with radius
$0.5''$. The admitted simplicity of this choice of source properties
should be irrelevant for demonstrating the feasibility of our method.

Starting from the known lensing potential of the cluster (projected
along the three axes of the simulation cube), weakly and strongly
lensed images are produced from the synthetic background sources by
tracing rays through the deflection-angle field, which is the gradient
of the potential.

\subsection{Results}

\subsubsection{Full knowledge of the critical curve}

\begin{figure}[ht]
  \includegraphics[width=\hsize]{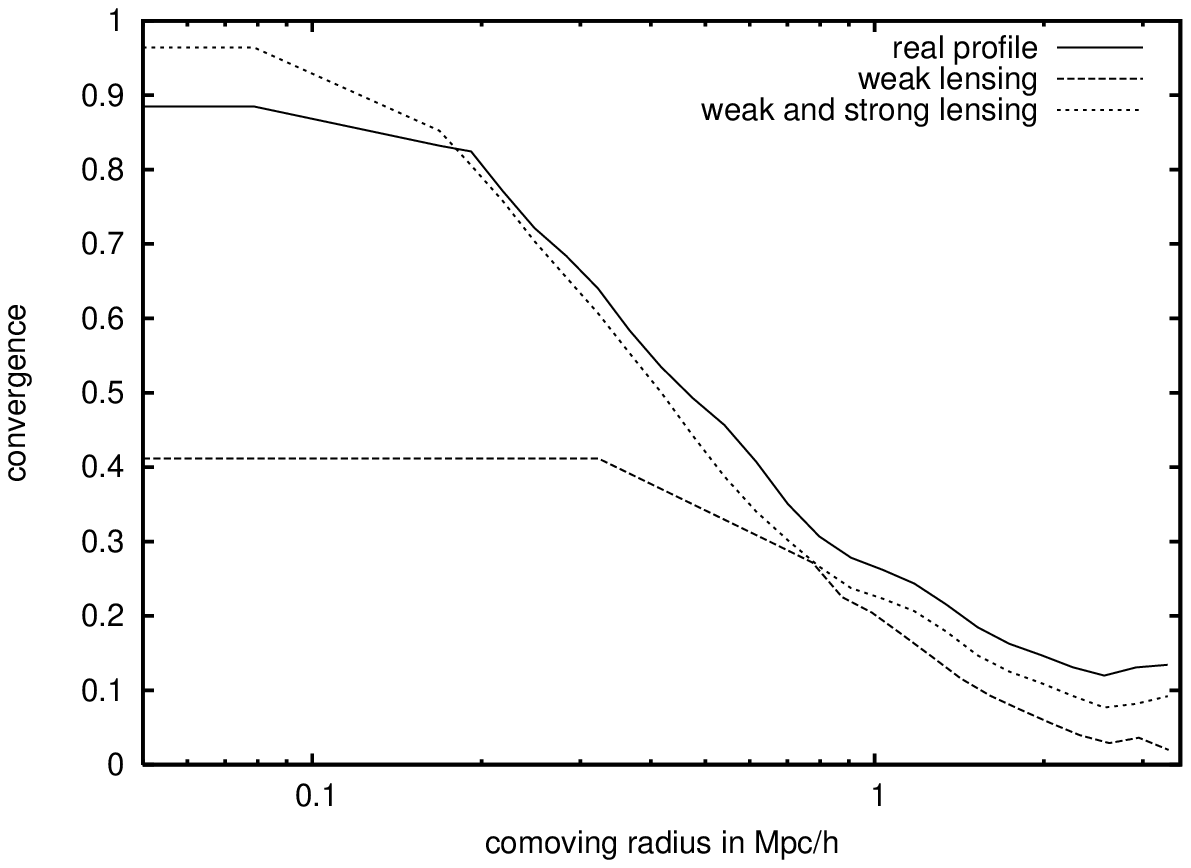}
  \includegraphics[width=\hsize]{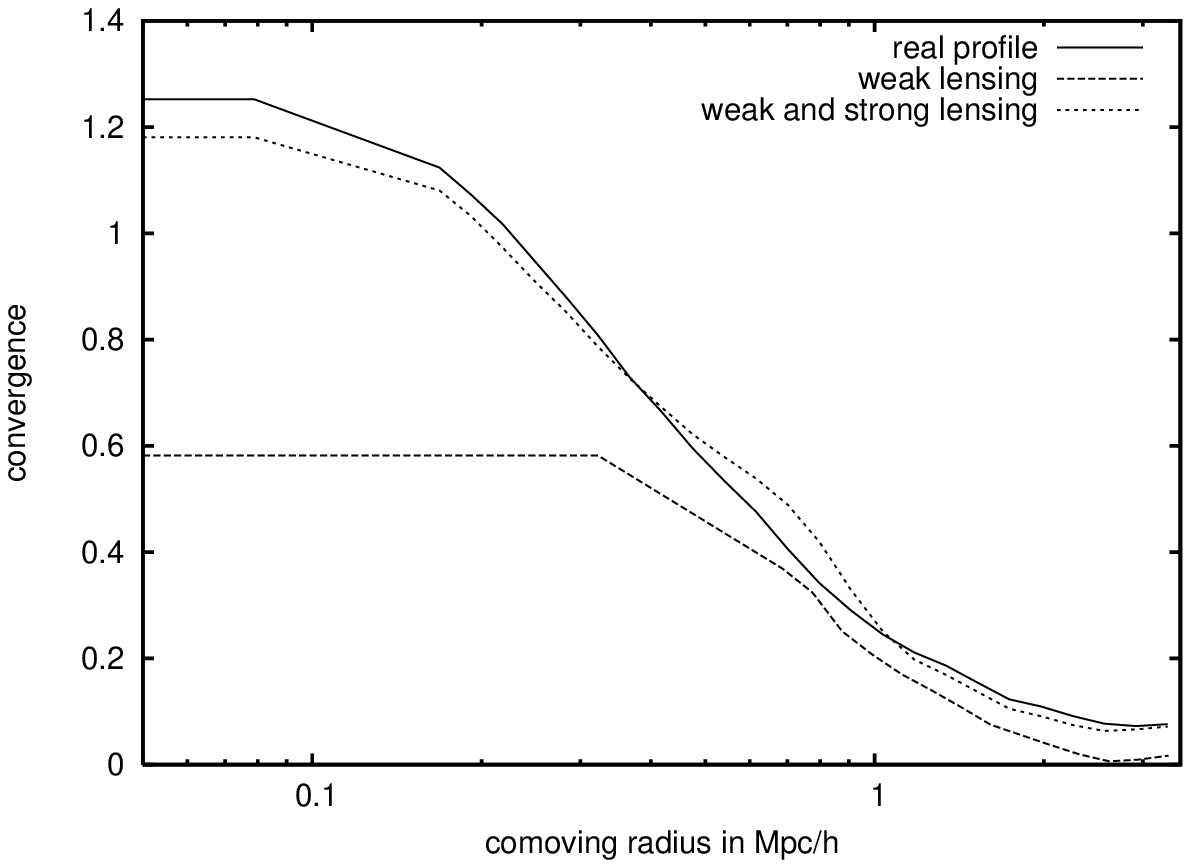}
\caption{Comparisons of the radial convergence profiles $\kappa(r)$
  for two projections of the original cluster, their weak-lensing
  reconstructions, and the joint reconstructions after adding the
  strong-lensing constraints. While the profile obtained from weak
  lensing alone suffers from the inevitable softening, strong lensing
  considerably improves the agreement.}
\label{fig:1}
\end{figure}

Figure~\ref{fig:1} shows the original and reconstructed radial
convergence profiles. Far away from the cluster centre, the
convergence is well recovered from weak-lensing data, but the central
part suffers from the inevitable softening due to the resolution limit
of weak lensing. This is corrected by adding the strong-lensing
constraints. The agreement between the outer parts of the true profile
and its weak-lensing reconstruction could be improved by means of the
transformation (\ref{eq:10}), but this would not remove the central
discrepancy.

\begin{figure}[ht]
  \includegraphics[width=\hsize]{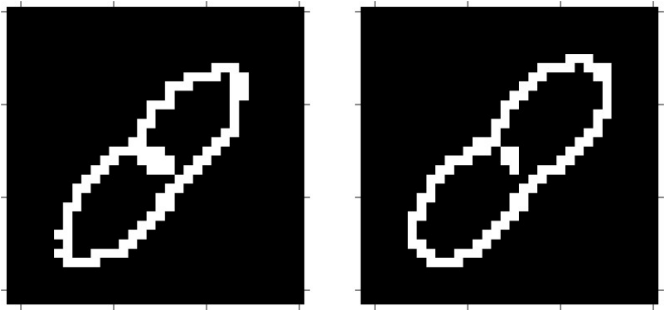}
\caption{\emph{Left panel:} original critical curve of a simulated
  cluster. \emph{Right panel:} critical curve reconstructed with the
  algorithm proposed by \cite{ME05.1}; see the text for details. Only
  the central part of the cluster is shown. The corresponding side
  length is $\approx1'$ with pixel size of $\approx2''$.}
\label{fig:critical_lines}
\end{figure}

The overall feasibility of the joint reconstruction is supported by
the agreement between the original critical lines and those
reconstructed using the algorithm by \cite{ME05.1}.
Figure~\ref{fig:critical_lines} shows the pixelised critical curve for
one of the simulated clusters used to test the method, and its
reconstruction. A direct comparison is possible because the simulation
also provides the location of the entire critical curve. At the
resolution used in our reconstruction algorithm, the agreement is
completely satisfactory. The chain of pixels covering the critical
curve is reproduced in great detail, providing ideal conditions for
the constraint from minimising $\chi^2_\mathrm{s}$.

\begin{figure}[ht]
  \includegraphics[width=0.49\hsize]{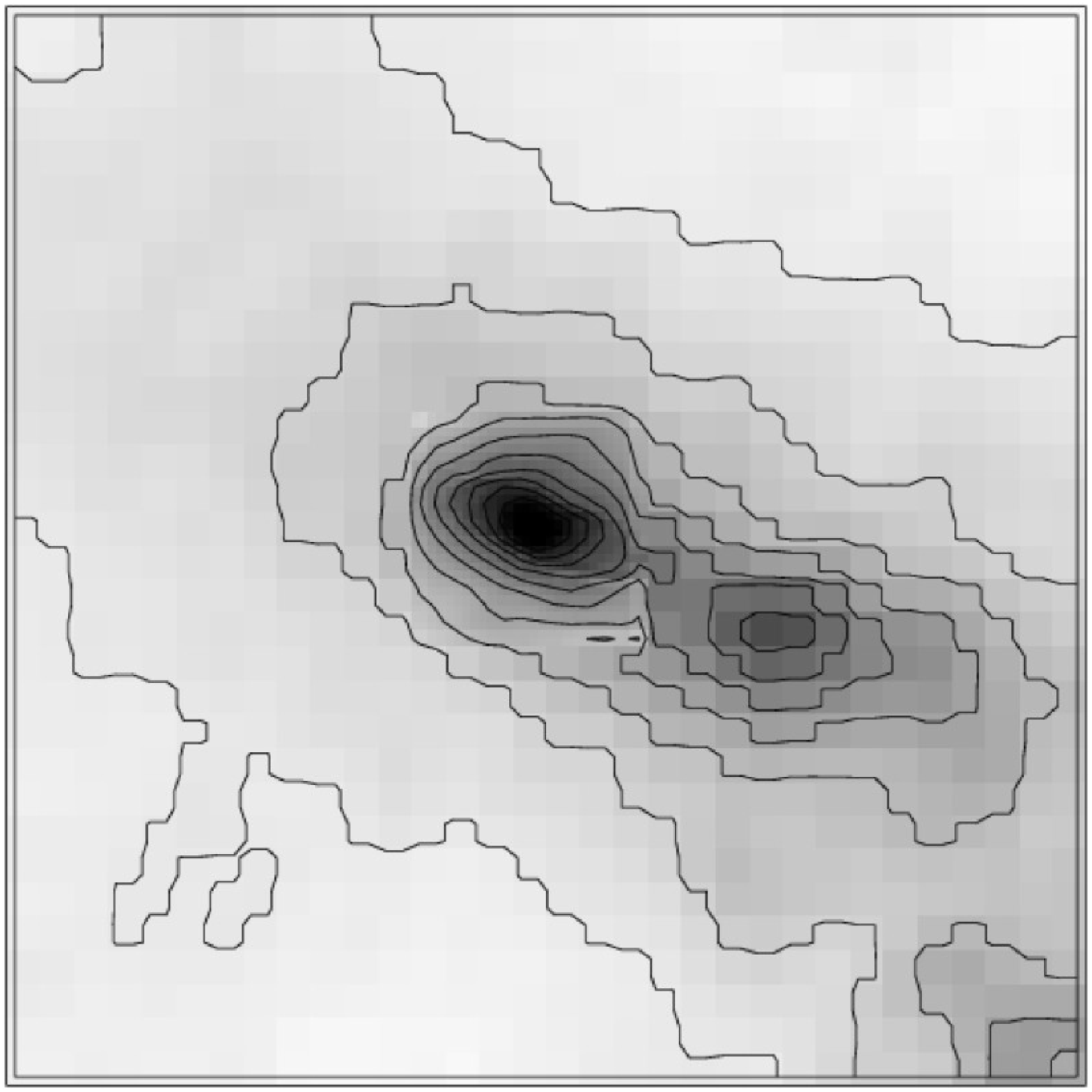}\hfill
  \includegraphics[width=0.49\hsize]{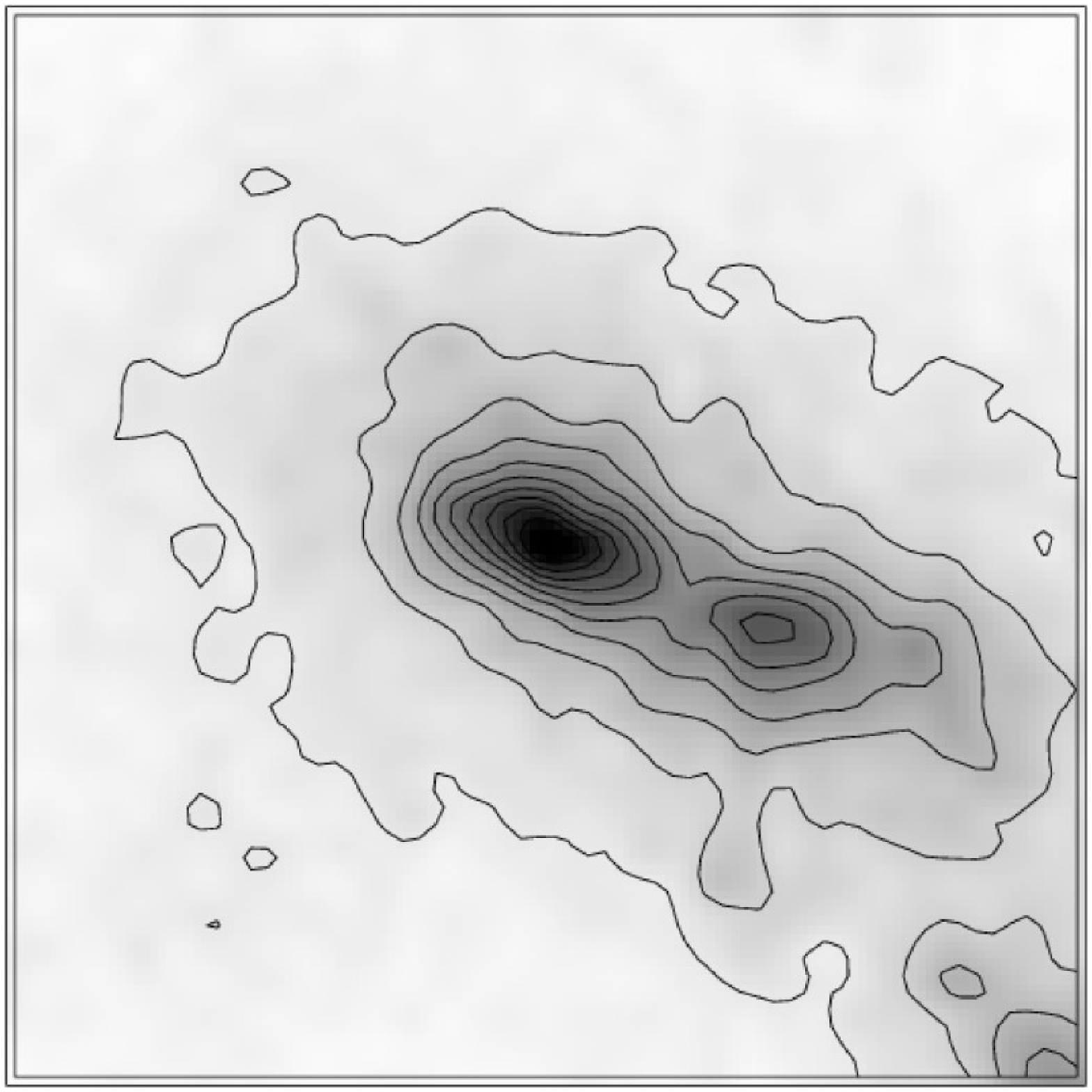}
  \includegraphics[width=0.49\hsize]{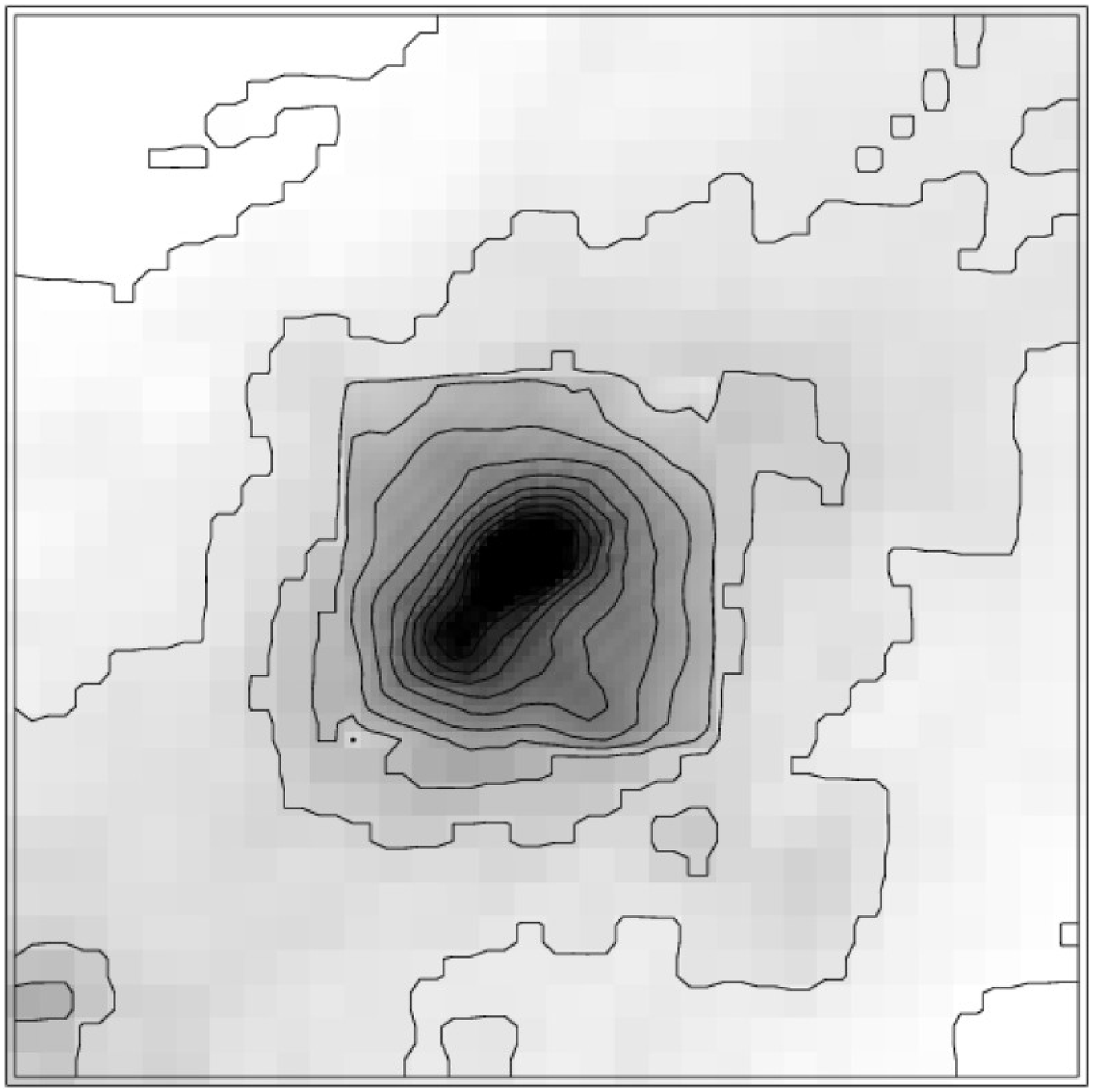}\hfill
  \includegraphics[width=0.49\hsize]{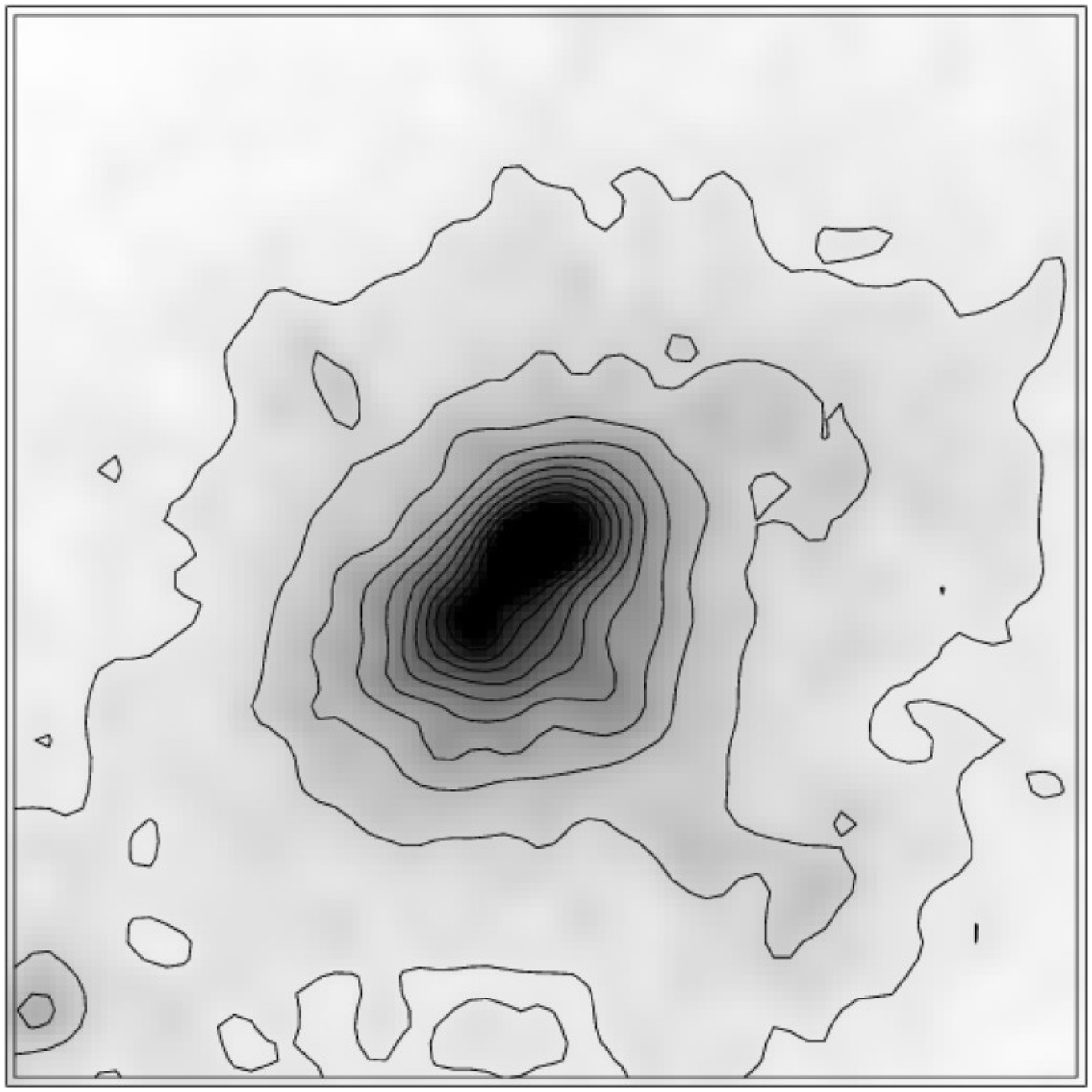}
\caption{Two simulated clusters are shown in the two rows. Their
  original convergence fields are displayed in the right column, their
  joint reconstructions using weak and strong lensing in the left. The
  colour scale is linear, the grey scales range from $0$ to $1.1$, and
  the contours are spaced by $\Delta\kappa=0.1$.}
\label{fig:2}
\end{figure}

Figure~\ref{fig:2} shows two-dimensional reconstructions of two
projections of the simulated cluster after applying the strong-lensing
correction in their cores. The coarse-grid resolution far from the
cluster centre is increased near the core where the critical curve of
the cluster is located. The resolution of the fine grid is 16 times
higher (per dimension) than that of the coarse grid, so that pixels of
the fine grid have $\sim2''$ side length.

\begin{figure}[ht]
  \includegraphics[width=\hsize]{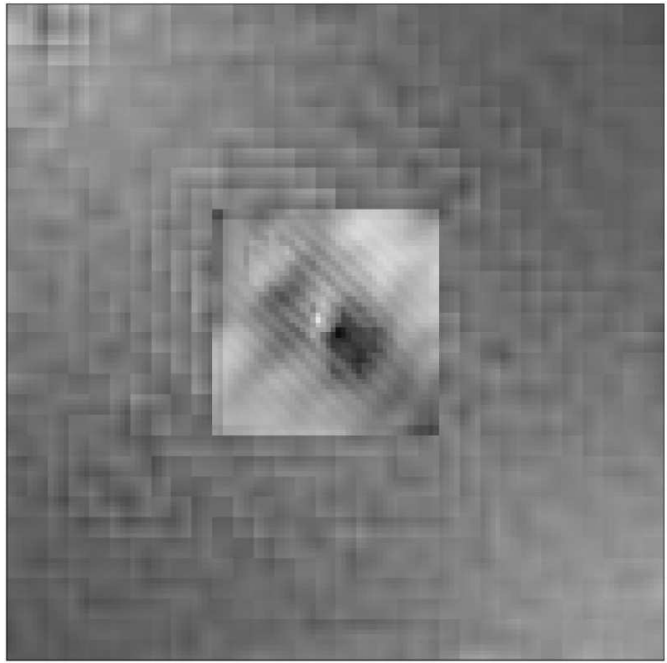}
\caption{Deviation of the reconstructed from the original convergence
  fields, $\kappa$ and $\bar\kappa$, respectively. To avoid
  divergences near the field boundaries where both $\kappa$ and
  $\bar\kappa$ are small, the grey scale encodes the relative
  difference between $(1+\kappa)$ and $(1+\bar\kappa)$,
  i.e.~$(\kappa-\bar\kappa)/(2+\kappa+\bar\kappa)$. The grey scale
  ranges from $-0.1$ (white) to $+0.1$ (black). Obviously, the
  relative deviations are very small.}
\label{fig:3}
\end{figure}

The quality of the combined reconstruction is also illustrated in
Fig.~\ref{fig:3} by a map of the relative differences between our
reconstruction and the original convergence of the simulated
cluster. Evidently, the mass distribution is well reproduced
everywhere in the field (see the caption for a quantitative
description), in agreement with the good recovery of the radial
profile. Moreover, constraints from strong lensing substantially
improve the reconstruction in the cluster core. This demonstrates that
the algorithm is working as expected.

\begin{figure}[ht]
  \includegraphics[width=0.49\hsize]{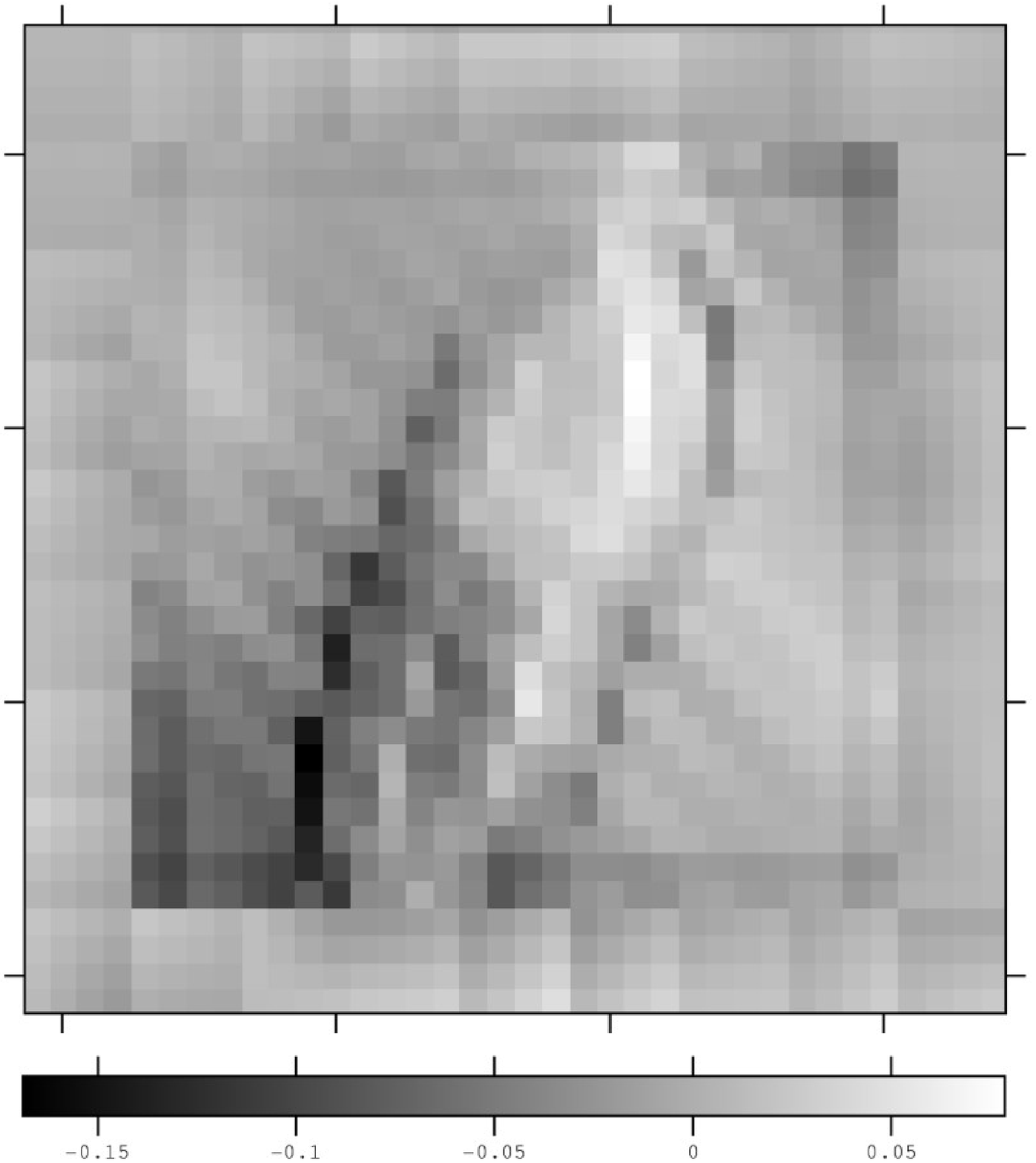}\hfill
  \includegraphics[width=0.49\hsize]{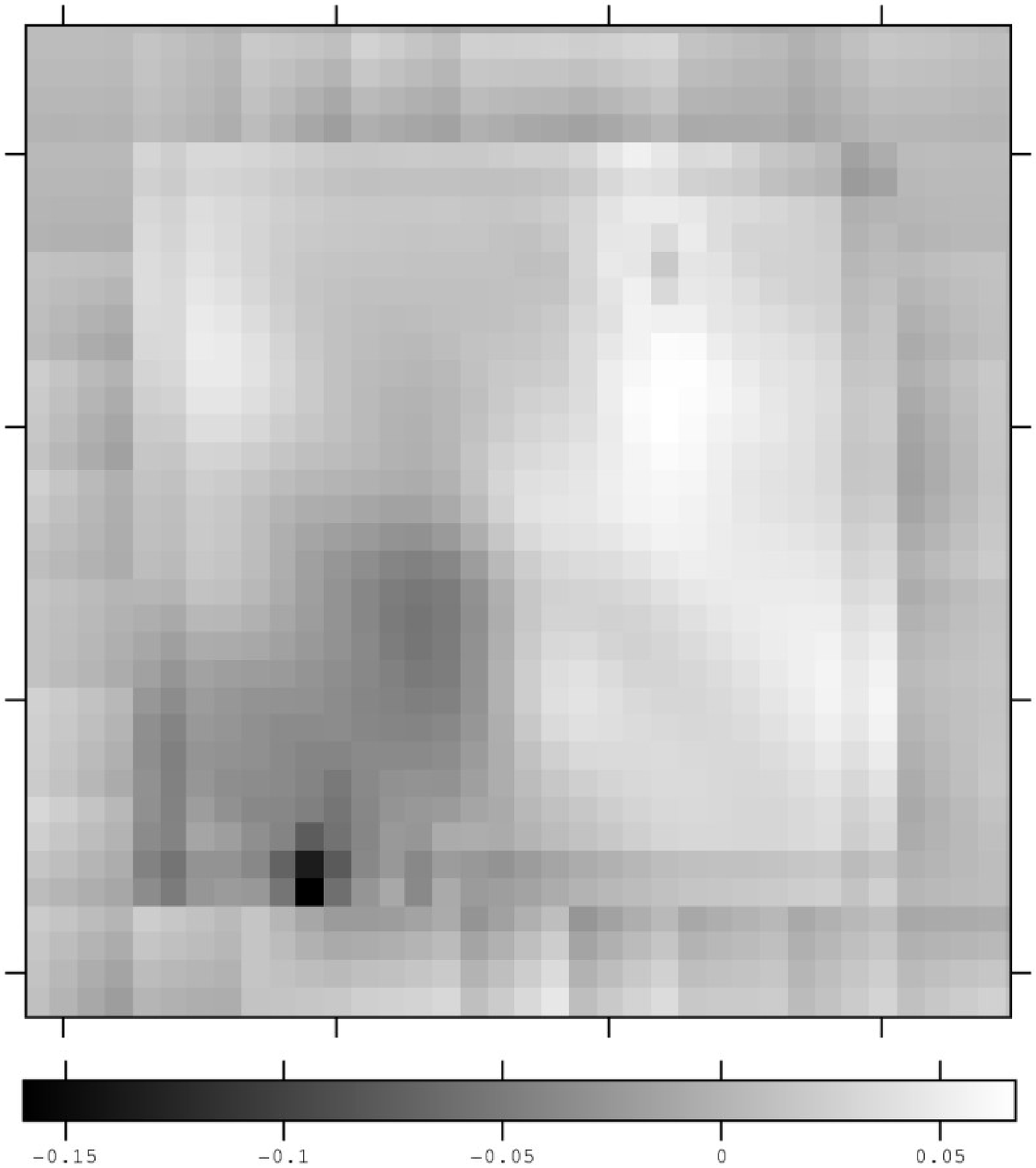}  
\caption{Both panels show the relative deviations between two
  reconstructed convergences $\kappa$ in the centre of a galaxy
  cluster. As in Fig.~\ref{fig:3}, the grey scale ranging from $-0.15$
  (black) to $+0.1$ (white) encodes
  $(\kappa-\bar\kappa)/(2+\kappa+\bar\kappa)$ to avoid divergences for
  $|\kappa|\approx0$. In both cases, the critical curve reconstructed
  with the method proposed by \cite{ME05.1} is used instead of the
  precise critical curve. \emph{Left panel:} Critical points are
  randomly shifted in the radial direction by one pixel or $\sim2''$
  prior to the reconstruction. While this procedure exaggerates
  realistic uncertainties, it provides an upper limit for the
  reconstruction errors which are $\lesssim10\%$ almost everywhere in
  the field. \emph{Right panel:} Only $16\%$ or 15 pixels of the
  critical curve are used, split into two sections opposing each other
  at the far ends of the critical curve. A random uncertainty in their
  positions is also included. Deviations are typically below
  $\sim10\%$.}
\label{fig:diff}
\end{figure}

Figure~\ref{fig:3} summarises the principal properties of our
reconstruction method. The outer region (with low resolution) is well
reproduced using only weak-lensing constraints, but the high-density
peak near the cluster core is poorly resolved. The reconstruction of
the core structure is greatly improved by adding the constraints from
strong lensing. Given the quality of the results, we did not need to
apply the iterative algorithm suggested above for correcting the
relation between measured ellipticities and shear.

\subsubsection{Approximate and partial knowledge of the critical
  curve}

Since our joint reconstruction algorithm rests on the approximate
knowledge of at least parts of the critical curve, we now investigate
two main sources of uncertainty, namely position errors in the
critical curve and incomplete knowledge of its location. For this
purpose, we run reconstructions taking these aspects into
account. Throughout, we do not assume precise knowledge of the
simulated cluster's critical curve, but reconstruct it using the
method by \cite[][see also \citealt{SA03.1}]{ME05.1}. Positional
errors of $\lesssim0.2''$ are found there. Since this is approximately
the spatial resolution of the fine grid on which we reconstruct the
convergence, we perform a reconstruction after randomly dislocating
critical points by one pixel each. This procedure yields an upper
limit to the effect of positional uncertainties in the critical
curves.

The left panel of Fig.~\ref{fig:diff} displays the relative difference
between the convergences reconstructed from the randomly shifted and
the original critical curves. The figure suggests that the algorithm
is working well because the inaccuracies in the critical curve cause
relative deviations at a typical level of a few per cent.

Addressing the second source of uncertainty, we assume that only part
of the critical curve is available (e.g.~from the positions and
orientations of one or more arcs). The right panel of
Fig.~\ref{fig:diff} shows the relative differences between one
reconstruction using the entire critical curve, and one using only
$16\%$ or 15~pixels of it. Mimicking real observations, we further
assume that this fraction of the critical curve is split into two
approximately equal sections at the far ends of the critical
curve. This approach simulates the presence of two arcs on opposite
sides of a cluster. The relative differences are remarkably low,
ranging around a few per cent and approaching $10\%$ only in few
points.

The above tests confirm the underlying idea that a reconstruction
based on joining weak and strong lensing features by using even part
of the critical curve is feasible. The main sources of uncertainty do
not substantially affect the method.

\section{Conclusions}

We have proposed a novel method for galaxy-cluster reconstruction
which combines weak and strong-lensing data. The method is based on a
least-$\chi^2$ fitting of the lensing potential $\psi$
\citep{BA96.3,SE98.4} and exploits the fact that the $\chi^2$
minimisation can be carried out efficiently by inverting a sparse
matrix \citep{BR05.2,BR05.1}. Contrary to other methods proposed for
joining weak and strong lensing information \citep{BR05.1,DI05.1}, we
propose to constrain the lensing potential obtained from the
weak-lensing data by the approximate location of the critical curves,
where the Jacobian determinant of the lens mapping must be close to
zero.

The angular resolution which can be achieved by weak-lensing cluster
reconstructions is typically of order $0.5'$ for background source
densities near $(30-40)\,\mathrm{arcmin^{-2}}$. This is much too
coarse for tracing critical curves, whose typical sizes are
$(0.5-1)'$. Thus, we propose to cover those regions of the cluster
fields with a refined grid where strong-lensing constraints are
available.

We test the performance of the method on synthetic images produced
with simulated lensing clusters. We conclude that the mass
distribution in galaxy clusters is well reproduced across the entire
field, in particular where constraints from strong lensing features
are introduced. In practice, our algorithm first finds a lensing
potential on the coarse grid which fits the weak-lensing data
best. This approximation to the potential is then interpolated into
the cells of the fine grid covering the critical curves or parts
thereof, and refined by a $\chi^2$ minimisation taking the
strong-lensing constraints into account.

We have used several approximations here. First, we treat weak lensing
to first order in the shear $\gamma$, i.e.~we compare measured
ellipticities to $\gamma$ rather than the reduced shear
$g=\gamma(1-\kappa)^{-1}$. Retaining the desirable linearity of the
method, this can easily be overcome by introducing an iteration scheme
in which the convergence $\kappa$ reconstructed in the previous step
is used to update the reduced shear of the current step (see also
\citealt{BR05.1}). Given the quality of our simulated results, we did
not need to use this iteration scheme.

Second, we have assumed knowledge of the entire critical curve, which
is of course not directly observable. Cluster images such as those
recently taken with ACS on-board HST, however, show so many large arcs
that critical curves can in fact be well constrained all around the
cluster cores. Furthermore, critical curves can reliably be inferred
from parametrised strong-lensing models, in particular when combined
with dynamical constraints from central cluster galaxies
(e.g.~\citealt{ME05.1}).

Third, we have assumed the strongly-lensed sources to be all at the
same redshift. If multiple arc systems at different redshifts are
observed, a lensing potential can still be reconstructed for one
fiducial source redshift $\bar z_\mathrm{s}$ by applying the distance
correction factors defined in (\ref{eq:02}).

Thus, we believe that the simplifications used here are not at all
restrictive, and that the method suggested here provides a useful
alternative to those proposed earlier. Our tests with randomly
displaced critical points and partial knowledge of critical curves
demonstrate that the method works well even in presence of positional
uncertainties and gaps in the known critical curve.

\acknowledgements{This work was supported by the Vigoni programme of
  the German Academic Exchange Service (DAAD) and Conference of
  Italian University Rectors (CRUI). We thank an anonymous referee for
  insightful comments.}

\appendix

\section{Linear equations for $\chi^2$ minimisation}

We summarise here some technical aspects of how the physical
quantities $\kappa$ and $\gamma$ are related to the discretised
deflection potential $\psi$ in our approach.

Derivatives of $\psi$ are replaced by common finite-differencing
schemes. We use 9 grid points for $\kappa$, 7 grid points for
$\gamma_2$ and 4 points for $\gamma_1$. This allows any lensing
quantity to be written as the multiplication of a well-defined matrix
with the vector of lensing-potential values,
\begin{equation}
  \kappa_i=\mathcal{K}_{ij}\psi_j\quad,\quad
  \gamma_i^1=\mathcal{G}_{ij}^1\psi_j\quad\hbox{and}\quad
  \gamma_i^2=\mathcal{G}_{ij}^2\psi_j\;.
\label{eq:A1}
\end{equation}
The matrices $\mathcal{G}^i$ and $\mathcal{K}$ are very sparse because
the finite-differencing schemes use only near neighbours
(cf.~\citealt{BR05.1}). Algorithms exist for efficient inversion of
such matrices.

Based on the finite-differencing schemes expressed by the matrix
Eqs..~(\ref{eq:A1}), the minimisation of $\chi^2$ is reduced to a
linear algebraic equation. Starting from the $\chi_\mathrm{w}^2$ for
weak lensing, we have
\begin{eqnarray}
0&=&\frac{\partial{\chi_w}^2(\psi_k)}{\partial\psi_j}\nonumber\\
&=&-2\sum_{i=1}^n\frac{1}{\sigma_{w i}^2}\left[
  (\epsilon_i^1-\gamma_i^1)\frac{\partial\gamma_i^1}{\partial\psi_j}+
  (\epsilon_i^2-\gamma_i^2)\frac{\partial\gamma_i^2}{\partial\psi_j}
  \right]\nonumber\\
&=&\sum_{i=1}^n\frac{-2}{\sigma_{w i}^2}\left[
    (\epsilon_i^1-\gamma_i^1)
    \frac{\partial\left(\mathcal{G}^1_{ik}\psi_k\right)}
         {\partial\psi_j}+
    (\epsilon_i^2-\gamma_i^2)
    \frac{\partial\left(\mathcal{G}^2_{ik}\psi_k\right)}
         {\partial\psi_j}
  \right]\nonumber\\
&=&\sum_{i=1}^n\frac{-2}{\sigma_{w i}^2}\left[
    (\epsilon_i^1-\gamma_i^1)\mathcal{G}^1_{ik}\delta_{jk}+
    (\epsilon_i^2-\gamma_i^2)\mathcal{G}^2_{ik}\delta_{jk}
  \right]\nonumber\\
&=&\sum_{i=1}^n\frac{-2}{\sigma_{w i}^2}\left[
    (\epsilon_i^1-\gamma_i^1)\mathcal{G}^1_{ij}+
    (\epsilon_i^2-\gamma_i^2)\mathcal{G}^2_{ij}
  \right]\nonumber\\
&=&\sum_{i=1}^n\frac{-2}{\sigma_{w i}^2}\Big[
    \epsilon_i^1\mathcal{G}^1_{ij}-
    \mathcal{G}^1_{ij}\mathcal{G}^1_{ik}\psi_k+
    \epsilon_i^2\mathcal{G}^2_{ij}-
    \mathcal{G}^2_{ij}\mathcal{G}^2_{ik}\psi_k
  \Big]\nonumber\\
&=&\sum_{i=1}^n\frac{2}{\sigma_{w i}^2}\left\{\left[
    \mathcal{G}^{1T}_{ji}\mathcal{G}^1_{ik}+
    \mathcal{G}^{2T}_{ji}\mathcal{G}^2_{ik}
  \right]\psi_k-\left[
    \epsilon_i^1\mathcal{G}^1_{ij}+\epsilon_i^2\mathcal{G}^2_{ij}
  \right]\right\}\;,
\label{eq:A2}
\end{eqnarray}
which can clearly be written in the form
\begin{equation}
  \mathcal{B}_{jk}\psi_k=\mathcal{V}_j\;,
\label{eq:A3}
\end{equation}
with the matrix
\begin{equation}
  \mathcal{B}_{jk}\equiv\sum_{i=1}^n\frac{1}{\sigma_{w i}^2}\left[
    \mathcal{G}_{ji}^{1T}\mathcal{G}_{ik}^1+
    \mathcal{G}_{ji}^{2T}\mathcal{G}_{ik}^2
  \right]
\label{eq:A4}
\end{equation}
and the data vector
\begin{equation}
  \mathcal{V}_j=\sum_{i=1}^n\frac{1}{\sigma_{w i}^2}\Big[
  \epsilon^1_i\mathcal{G}_{ij}^1+\epsilon^2_i\mathcal{G}_{ij}^2
  \Big]\;.
\label{eq:A5}
\end{equation}

Similarly evaluating the constraints from strong lensing yields
\begin{eqnarray}
\frac{\partial\chi_\mathrm{s}^2}{\partial\psi^*_{j^\prime}}
 &=&\frac{\partial}{\partial\psi^*_{j^\prime}}\left[
      \sum_{i=1}^{n*}\frac{(\det\mathcal{A}_i)^2}{\sigma^2_i}
    \right]\nonumber\\
 &=&\sum_{i=1}^{n*}\frac{-4\det\mathcal{A}_i}{\sigma^2_i}\left[
      (1-\kappa_i)\frac{\partial\kappa_i}{\partial\psi^*_{j^\prime}}-
      \frac{\partial\gamma_i^1}{\partial\psi^*_{j^\prime}}-
      \frac{\partial\gamma_i^2}{\partial\psi^*_{j^\prime}}
    \right]\nonumber\\
 &=&\sum_{i=1}^{n*}\frac{-4\det\mathcal{A}_i}{\sigma^2_i}\left[
      (1-\kappa_i)\mathcal{K}_{ij^\prime}-
      \gamma_i^1\mathcal{G}^1_{ij^\prime}-
      \gamma_i^2\mathcal{G}^2_{ij^\prime}
    \right]\nonumber\\
 &=&\mathcal{T}_{j^\prime}\;,
\label{eq:A6}
\end{eqnarray}
where the $\kappa_i$ and the $\gamma_i^{1,2}$ are obtained from the
interpolated weak-lensing solution.

On the refined grid, the $\chi^2$ minimisation (\ref{eq:A3}) is
modified by
\begin{equation}
  \mathcal{B}_{jk}\psi_k^*=\mathcal{V}_j-\mathcal{T}_{j^\prime}\;.
\label{eq:A7}
\end{equation}
Obviously, if the weak-lensing solution already satisfies
$\det\mathcal{A}=0$ on the critical curves, $\mathcal{T}=0$ there, and
no correction is necessary.

\bibliography{master}

\begin{thebibliography}{18}
\expandafter\ifx\csname natexlab\endcsname\relax\def\natexlab#1{#1}\fi

\bibitem[{Bartelmann {et~al.}(1998)Bartelmann, Huss, Colberg, Jenkins, \&
  Pearce}]{BA98.2}
Bartelmann, M., Huss, A., Colberg, J.~M., Jenkins, A., \& Pearce, F.~R. 1998,
  A\&A, 330, 1

\bibitem[{Bartelmann {et~al.}(1996)Bartelmann, Narayan, Seitz, \&
  Schneider}]{BA96.3}
Bartelmann, M., Narayan, R., Seitz, S., \& Schneider, P. 1996, ApJL, 464, L115

\bibitem[{{Brada{\v c}} {et~al.}(2005{\natexlab{a}}){Brada{\v c}}, {Erben},
  {Schneider}, {Hildebrandt}, {Lombardi}, {Schirmer}, {Miralles}, {Clowe}, \&
  {Schindler}}]{BR05.2}
{Brada{\v c}}, M., {Erben}, T., {Schneider}, P., {et~al.} 2005{\natexlab{a}},
  \aap, 437, 49

\bibitem[{{Brada{\v c}} {et~al.}(2005{\natexlab{b}}){Brada{\v c}}, {Schneider},
  {Lombardi}, \& {Erben}}]{BR05.1}
{Brada{\v c}}, M., {Schneider}, P., {Lombardi}, M., \& {Erben}, T.
  2005{\natexlab{b}}, \aap, 437, 39

\bibitem[{{Brainerd} {et~al.}(1996){Brainerd}, {Blandford}, \&
  {Smail}}]{BR96.1}
{Brainerd}, T.~G., {Blandford}, R.~D., \& {Smail}, I. 1996, ApJ, 466, 623

\bibitem[{{Broadhurst} {et~al.}(2005){Broadhurst}, {Ben{\'{\i}}tez}, {Coe},
  {Sharon}, {Zekser}, {White}, {Ford}, {Bouwens}, {Blakeslee}, {Clampin},
  {Cross}, {Franx}, {Frye}, {Hartig}, {Illingworth}, {Infante}, {Menanteau},
  {Meurer}, {Postman}, {Ardila}, {Bartko}, {Brown}, {Burrows}, {Cheng},
  {Feldman}, {Golimowski}, {Goto}, {Gronwall}, {Herranz}, {Holden}, {Homeier},
  {Krist}, {Lesser}, {Martel}, {Miley}, {Rosati}, {Sirianni}, {Sparks},
  {Steindling}, {Tran}, {Tsvetanov}, \& {Zheng}}]{BR05.4}
{Broadhurst}, T., {Ben{\'{\i}}tez}, N., {Coe}, D., {et~al.} 2005, ApJ, 621, 53

\bibitem[{{Diego} {et~al.}(2005){Diego}, {Tegmark}, {Protopapas}, \&
  {Sandvik}}]{DI05.1}
{Diego}, J.~M., {Tegmark}, M., {Protopapas}, P., \& {Sandvik}, H.~B. 2005,
  ArXiv Astrophysics e-prints, arXiv:astro-ph/0509103

\bibitem[{Falco {et~al.}(1985)Falco, Gorenstein, \& Shapiro}]{FA85.1}
Falco, E.~E., Gorenstein, M.~V., \& Shapiro, I.~I. 1985, ApJ, 289, L1

\bibitem[{{Ford} {et~al.}(1996){Ford}, {Feldman}, {Golimowski}, {Tsvetanov},
  {Bartko}, {Crocker}, {Bely}, {Brown}, {Burrows}, {Clampin}, {Hartig},
  {Postman}, {Rafal}, {Sparks}, {White}, {Broadhurst}, {Illingworth}, {Kelly},
  {Woodruff}, {Cheng}, {Kimble}, {Krebs}, {Neff}, {Lesser}, \&
  {Miley}}]{FO96.1}
{Ford}, H.~C., {Feldman}, P.~D., {Golimowski}, D.~A., {et~al.} 1996, in Proc.
  SPIE Vol. 2807, p. 184-196, Space Telescopes and Instruments IV, Pierre Y.
  Bely; James B. Breckinridge; Eds., ed. P.~Y. {Bely} \& J.~B. {Breckinridge},
  184--196

\bibitem[{{Grossman} \& {Narayan}(1988)}]{GR88.1}
{Grossman}, S.~A. \& {Narayan}, R. 1988, ApJL, 324, L37

\bibitem[{{Grossman} \& {Narayan}(1989)}]{GR89.1}
{Grossman}, S.~A. \& {Narayan}, R. 1989, ApJ, 344, 637

\bibitem[{{Kovner}(1989)}]{KO89.1}
{Kovner}, I. 1989, ApJ, 337, 621

\bibitem[{Meneghetti {et~al.}(2005)Meneghetti, Bartelmann, Jenkins, \&
  Frenk}]{ME05.1}
Meneghetti, M., Bartelmann, M., Jenkins, A., \& Frenk, C. 2005, ArXiv
  Astrophysics e-prints; arXiv:astro-ph/0509323

\bibitem[{{Rix} {et~al.}(2004){Rix}, {Barden}, {Beckwith}, {Bell}, {Borch},
  {Caldwell}, {H{\"a}ussler}, {Jahnke}, {Jogee}, {McIntosh}, {Meisenheimer},
  {Peng}, {Sanchez}, {Somerville}, {Wisotzki}, \& {Wolf}}]{RI04.2}
{Rix}, H.-W., {Barden}, M., {Beckwith}, S.~V.~W., {et~al.} 2004, ApJS, 152, 163

\bibitem[{Sand {et~al.}(2004)Sand, Treu, Smith, \& Ellis}]{SA03.1}
Sand, D.~J., Treu, T., Smith, G.~P., \& Ellis, R.~S. 2004, ApJ, 604, 88

\bibitem[{Schneider {et~al.}(1992)Schneider, Ehlers, \& Falco}]{SC92.1}
Schneider, P., Ehlers, J., \& Falco, E.~E. 1992, Gravitational Lenses (Springer
  Verlag, Heidelberg)

\bibitem[{{Schneider} \& {Seitz}(1995)}]{SC95.1}
{Schneider}, P. \& {Seitz}, C. 1995, A\&A, 294, 411

\bibitem[{Seitz {et~al.}(1998)Seitz, Schneider, \& Bartelmann}]{SE98.4}
Seitz, S., Schneider, P., \& Bartelmann, M. 1998, A\&A, 337, 325

\end{thebibliography}
\bibliographystyle{aa}

\end{document}